\documentclass[pra,amsmath,amssymb,reprint,superscriptaddress]{revtex4-1}

\usepackage{graphicx}
\usepackage{dcolumn}
\usepackage{bm}

\usepackage[utf8]{inputenc}
\usepackage[T1]{fontenc}
\usepackage{etoolbox}

\usepackage{txfonts}

\newcommand{\p}{\mathrm{p}}
\renewcommand{\i}{\mathrm{i}}
\newcommand{\s}{\mathrm{s}}
\newcommand{\sinc}{\mathrm{sinc}}
\newcommand{\e}{\mathrm{e}}
\newcommand{\vac}{\text{vac}}
\newcommand{\twin}{\text{twin}}

\newcommand{\SE}{\mathrm{SE}}
\newcommand{\GSB}{\mathrm{GSB}}
\newcommand{\ESA}{\mathrm{ESA}}
\newcommand{\rephasing}{\mathrm{r}}
\newcommand{\nonrephasing}{\mathrm{nr}}

\newcommand{\dt}{\Delta t}
\newcommand{\adagger}{\hat{a}^\dagger}

\renewcommand{\Im}{\mathrm{Im}}
\renewcommand{\Re}{\mathrm{Re}}

\newcommand{\E}{\hat{E}}

\makeatletter
\def\@email#1#2{%
 \endgroup
 \patchcmd{\titleblock@produce}
  {\frontmatter@RRAPformat}
  {\frontmatter@RRAPformat{\produce@RRAP{*#1\href{mailto:#2}{#2}}}\frontmatter@RRAPformat}
  {}{}
}%
\makeatother
\begin{document}


\title{Pathway selectivity in time-resolved spectroscopy using two-photon coincidence counting with quantum entangled photons}
\author{Yuta Fujihashi}
\affiliation{Department of Engineering Science, The University of Electro-Communications, Chofu 182-8585, Japan}
\email{fujihashi@uec.ac.jp}
\author{Akihito Ishizaki}
\affiliation{Institute for Molecular Science, National Institutes of Natural Sciences, Okazaki 444-8585, Japan}
\affiliation{Graduate Institute for Advanced Studies, SOKENDAI, Okazaki 444-8585, Japan}
\author{Ryosuke Shimizu}
\affiliation{Department of Engineering Science, The University of Electro-Communications, Chofu 182-8585, Japan}
\affiliation{Institute for Advanced Science, The University of Electro-Communications, Chofu 182-8585, Japan}


\begin{abstract}
Ultrafast optical spectroscopy is a powerful technique for studying the dynamic processes of molecular systems in condensed phases. However, in molecular systems containing many dye molecules, the spectra can become crowded and difficult to interpret owing to the presence of multiple nonlinear optical contributions. 
In this work, we theoretically propose time-resolved spectroscopy based on the coincidence counting of two entangled photons generated via parametric down-conversion with a monochromatic laser. We demonstrate that the use of two-photon counting detection of entangled photon pairs enables the selective elimination of the excited-state absorption signal. This selective elimination cannot be realized with classical coherent light.
We anticipate that the proposed spectroscopy will help to simplify the spectral interpretation in complex molecular and materials systems comprising multiple molecules.
\end{abstract}

\maketitle

%


\section{Introduction}
Quantum light is a state of light that exhibits unique properties in photon statistics and correlations that are not found in classical light. By harnessing the quantum nature of light, it is possible to achieve selective excitation and precise measurements that cannot be performed in spectroscopic measurements using classical light such as lasers \cite{Yabushita:2004hy, Oka:2010if, Schlawin:2013dq, Kalashnikov:2016cl, Mukai:2021qu, Eto:2021en, Matsuzaki:2022su, Albarelli:2023fu, Khan:2023do}. 
The use of quantum light for spectroscopic measurements can be traced back to the field of atomic, molecular, and optical physics \cite{Gea1989:tw, Javanainen1990:li, Georgiades:1995dd, Georgiades:1997at, Fei:1997es, Saleh:1998vl}. 
With advances in quantum optical technology, there was increased interest in the use of quantum light in nonlinear spectroscopic measurements of condensed phase molecular systems \cite{Lee:2006id, Roslyak:2009cy, Upton:2013is, Dorfman:2014bn, Dorfman:2016da, Schlawin:2016er, Debnath:2020en, Dorfman:2021ho, Asban:2021in, Asban:2021di, Asban:2022no, Schlawin:2022po, Zhang:2022en, Kizmann:2023qu, Ko:2023em}.
For example, by utilizing non-classical correlations between entangled photons, time-resolved spectroscopy using only a monochromatic laser has been proposed \cite{Ishizaki:2020jl, Chen:2022mo, Gu:2023ph, Fujihashi:2023pr}.
It has also been experimentally reported that the Hong–Ou–Mandel interferometer with entangled photons enables the measurement of molecular dephasing times at the femtosecond time scale without the need for ultrashort laser pulses \cite{Kalashnikov:2017hx, Eshun:2021in}.

Theoretically, the nonlinear optical response obtained by irradiating light onto a material is expressed as the convolution of the response function of the material and the multi-body correlation function of the electric field \cite{Mukamel:1995us, Dorfman:2016da}. In the case of classical light, the correlation function of an electric field can be expressed as a product of the laser pulse envelopes. In conventional nonlinear spectroscopy, such as the pump-probe and photon-echo technique, the third-order nonlinear optical response induced by multiple laser pulses on a material includes stimulated emission (SE), ground-state bleaching (GSB), and excited-state absorption (ESA) signals. The GSB and SE signals involve the electronic ground state and single-excitation manifold, while the ESA signal involves their electronic states plus the double-excitation manifold.
In the absence of electronic coupling between two molecules, the ESA peak that originates from the electronic state of a molecule exactly cancels out the corresponding GSB peak because the signs of ESA and GSB are opposite.
However, in the presence of electronic coupling, the ESA signal shifts slightly from the peak positions of the GSB, and does not cancel with the GSB \cite{SchlauCohen:2011cw}. 
This results in a spectrum with many peaks. Moreover, the broadening of the spectrum leads to overlapping of these peak contributions, creating a congested spectral profile. 
This complexity is particularly pronounced in biomolecular processes, such as photosynthetic light-capture systems containing many pigments.
In contrast, for quantum light, the correlation function of the radiation field is replaced by the expectation value of the product of the photon creation and annihilation operators.
The operators act on the quantum state of the light, where the exchange of operators is constrained by the commutation relation.
When a product of more annihilation operators than the number of photons presents in the quantum state of light acts on it, the expectation value is zero.
Owing to this property, only quantum correlation functions with a particular ordered sequence of the operators have finite values, thus limiting the types of Liouville pathways that contribute to the signal.
Indeed, it was reported that the use of both this non-classical photon correlation and quantum interferometry can enable the selective extraction of specific signal pathways in nonlinear spectra \cite{Asban:2021in, Asban:2021di, Asban:2022no, Kizmann:2023qu}.

Inspired by the path-selectivity provided by the quantum light and interferometry described above, in this study, we theoretically explore the potential of time-resolved spectroscopy based on the two-photon coincidence detection of entangled photon pairs generated via parametric down-conversion (PDC) pumped with a monochromatic laser. We demonstrate that by utilizing two-photon counting detection and the non-classical correlation of entangled photon pairs, the ESA signal pathway in the time-resolved spectra can be eliminated, and the SE signal can be selectively extracted.

\bigskip

\section{Experiment setup}

Before presenting a detailed theoretical description, we briefly discuss how the proposed two-photon coincidence entangled two-photon spectroscopy differs from the entangled two-photon spectroscopy in Ref.~\citenum{Ishizaki:2020jl}.  

In the entangled two-photon spectroscopy, the signal and idler photons are employed as pump and probe fields, respectively, with a delay interval.
The probe field transmitted through the sample is frequency-dispersed, and the change in the transmitted photon number is measured.
The signal is expressed as the convolution of the third-order response function of a molecule and the four-body correlation function of the electric field.
The signal corresponds to the spectral information along anti-diagonal lines of two-dimensional (2D) Fourier-transformed photon-echo spectra, which is given by the sum of the GSB, SE, and ESA signals.

\begin{figure}
    \centering
    \includegraphics{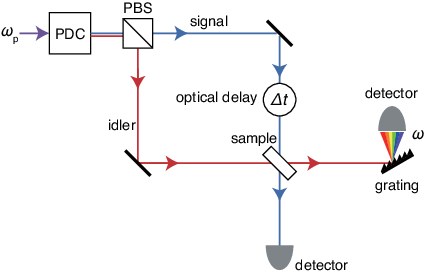}
    \caption{Schematic of the two-photon coincidence counting measurement with quantum entangled photon pairs generated via parametric down-conversion (PDC) pumped with a monochromatic laser of frequency $\omega_\p$.
    The delay interval is controlled by adjusting the path difference between the signal and idler beams.
    The signal photon is not resolved spectrally, whereas the idler photon transmitted through the sample is frequency-dispersed. 
    Therefore, changes in the two-photon counting rate with frequency $\omega$ are measured.}
    \label{fig1}
\end{figure}

In contrast, in the two-photon coincidence entangled two-photon spectroscopy presented in Fig.~\ref{fig1}, the measurement involves not only frequency-dispersing the probe field after it passes through the sample, but also detecting the transmitted pump field without frequency resolution. 
Thus, the signal is given by changes in the rate of two-photon counting, and the correlation function of the field operators in the signal is characterized as a six-body correlation function.
The addition of two photon operators restricts the combinations of the order of products of the operators for which the correlation function of the field operators is nonzero. As shown later, the resulting correlation function of the ESA signal vanishes.

\section{Theory}
We consider a system consisting of the light field and a molecular system. The total Hamiltonian is given by $\hat{H} = \hat{H}_{\rm mol} + \hat{H}_{\rm field} + \hat{H}_{\rm mol-field}$. The first term, $\hat{H}_{\rm mol}$, gives the Hamiltonian of the photoactive degrees of freedom (DOFs) in molecules. The second term describes the free electric field. In this work, the electronic ground state $\lvert 0 \rangle$, single-excitation manifold $\{\lvert e_\alpha \rangle\}$, and double-excitation manifold $\{\lvert f_{\bar\gamma} \rangle\}$ are considered as photoactive DOFs. 
The overline of the subscripts indicates the state in the double-excitation manifold. 
The optical transitions are described by the dipole operator $\hat{\mu} = \hat{\mu}_+ + \hat{\mu}_-$, where $\hat{\mu}_- = \sum_\alpha \mu_{\alpha 0} \lvert 0 \rangle \langle e_\alpha \rvert + \sum_{\alpha \bar\gamma} \mu_{\bar\gamma \alpha} \lvert e_\alpha \rangle\langle f_{\bar\gamma} \rvert$ and $\hat{\mu}_+ = \hat{\mu}_-^\dagger$. 
The rotating-wave approximation enables the expression of the molecule-field interaction as $\hat{H}_{\rm mol-field}(t) = -\hat{\mu}_- \E^+(t) -\hat{\mu}_+ \E^-(t)$, where $\E^+(t)$ and $\E^-(t)$ denote the positive- and negative-frequency components of the electric field operator, respectively.

We consider the setup shown in Fig.~\ref{fig1}.
A pair of entangled photons created by PDC pumped with a monochromatic laser of frequency $\omega_\p$ are split using a polarized beam splitter. 
Although the relative delay between the signal and idler photons is innately determined by the entanglement time $T_\e$ \cite{Saleh:1998vl}, the delay interval is further controlled by adjusting the path difference between the beams. 
This controllable delay is denoted by $\dt$ in this study. In this measurement, the signal photon is employed as the pump field with time delay $\dt \ge 0$, whereas the idler photon is used for the probe field; hence, the positive-frequency component of the field operator is expressed as 
\begin{align}
    \E^+(t)
    =
    \E_\s^+(t)
    +
    \E_\i^+(t+\dt),    
\end{align}
with $\E_\sigma^+(t)=(2\pi)^{-1}\int d\omega\,\hat{a}_\sigma(\omega)e^{-i\omega t}$, where $\hat{a}_\s(\omega)$ and $\hat{a}_\i(\omega)$ are respectively the annihilation operator of the signal and idler photons at frequency $\omega$.
We adopt the slowly varying envelope approximation, in which the bandwidth of the field is assumed to have a negligible width compared to the central frequency \cite{loudon2000quantum}.
The pump field transmitted through sample $\E_\s$ is not resolved spectrally, whereas the transmitted probe field $\E_\i$ is frequency-dispersed. 
The signal is given by changes in the two-photon counting rate,
$\int_{-\infty}^\infty dt\, \mathrm{tr}[\E_\s^-(t)\E_\i^-(\omega)\E_\i^+(\omega)\E_\s^+(t)\hat{\rho}(\infty)]$, where the density operator $\hat{\rho}(t)$ represents the state of the total system.
The change of the two-photon counting signal can be recast as
\begin{align}
    S_\mathrm{TPC}(\omega;\dt)
    &=
    \Im
    \int_{-\infty}^\infty d\tau
    \int_{-\infty}^\infty dt \,
    e^{i\omega(t+\dt)}
\notag \\
    &\quad\times    
    \mathrm{tr}
    [\E_\s^-(\tau)\E_\s^+(\tau)\E_\i^-(\omega)\hat{\mu}_{-}\hat{\rho}(t)]
\notag \\
    &\quad
    +
    \Im 
    \int_{-\infty}^\infty dt \,
    \mathrm{tr}
    [\E_\i^-(\omega)\E_\i^+(\omega)\E_\s^-(t)\hat{\mu}_{-}\hat{\rho}(t)]
    \label{eq:tpc-signal}
\end{align}
with the initial condition of $\hat{\rho}(-\infty)=\lvert 0 \rangle\langle 0 \rvert \otimes \lvert \psi_\twin \rangle\langle \psi_\twin \rvert$, where $\lvert \psi_\twin \rangle=\int d\omega_1 \int d\omega_2\,f(\omega_1,\omega_2) \adagger_\s(\omega_1) \adagger_\i(\omega_2)\vert \vac \rangle$ is specified in Appendix~A.
We can calculate Eq.~\eqref{eq:tpc-signal} by the perturbative expansion of $\hat{\rho}(t)$ with respect to the molecule-field interaction $\hat{H}_{\rm mol-field}$.
The resultant signal is expressed as the sum of eight contributions, which are classified into ground-state bleaching (GSB), stimulated emission (SE), excited-state absorption (ESA), and double-quantum coherence (DQC). Typically, the coherence between the electronic ground state and a doubly excited state rapidly decays in comparison to the others \cite{Fujihashi:2023pr}. Hence, the DQC contribution is disregarded in this work.
Consequently, Eq.~\eqref{eq:tpc-signal} can be expressed as
\begin{align}
    S_\mathrm{TPC}(\omega;\dt)
    =
    \Im \,
    \sum_{x,y}
    S_{x}^{(y)}(\omega;\dt),
\end{align}
with
\begin{align}
    S_x^{(y)}(\omega;\dt)
    &=
    \iiint_0^\infty d^3 s \,
    \Phi_{x}^{(y)}(s_3,s_2,s_1)
    \int_{-\infty}^\infty dt
\notag \\
    &\quad\times
    \Biggl[
    e^{i\omega(t+\dt)}
    \int_{-\infty}^\infty d\tau
     \,
    C_{x,\i\s}^{(y)}(\omega,t,\tau;s_3,s_2,s_1)
\notag \\
    &\quad      
    +
    C_{x,\s\i}^{(y)}(\omega,t,t;s_3,s_2,s_1)
    \Biggr],
    \label{eq:tpc-signal-contribution}
\end{align}
where $x$ indicates GSB, SE, or ESA, and $y$ indicates “rephasing” (r) or “non-rephasing” (nr).
Here, $\Phi_{x}^{(y)}(s_3,s_2,s_1)$ is the third-order response function of the molecule, whereas $C_{x,\sigma\sigma'}^{(y)}(\omega,t,\tau;s_3,s_2,s_1)$ is the six-body correlation function of the field operators, such as $C_{\SE,\sigma\sigma'}^{(\rephasing)}(\omega,t,\tau;s_3,s_2,s_1)=\langle\E^-(t-s_3-s_2-s_1)\E^+(t-s_3)\E_{\sigma'}^-(\tau)\E_{\sigma'}^+(\tau)\E_{\sigma}^-(\omega)\E^+(t-s_3-s_2)\rangle$.
The bracket indicates the expectation value in terms of $\lvert\psi_\twin \rangle$, namely, $\langle \psi_\twin \rvert \cdots \lvert\psi_\twin \rangle$.

In the following, we focus on the rephasing SE and ESA contributions for illustrative purposes. Further details on the other signal contributions can be found in Appendix~B.

\subsection{Rephasing SE contribution}

To obtain a concrete expression of the signal, the memory effect, the memory effect straddling different time intervals in the response function is ignored \cite{Ishizaki:2008be}. 
Consequently, the response function is expressed as \cite{Ishizaki:2012kf}
\begin{align}
    \Phi_{\SE}^{(\rephasing)}(t_3,t_2,t_1)
    &=
    \left( \frac{i}{\hbar} \right)^3
    \sum_{\alpha\beta\gamma\delta}
    \mu_{\delta0}
    \mu_{\gamma0}
    \mu_{\beta0}
    \mu_{\alpha0}
\notag \\
    &\quad\times     
    G_{\gamma0}(t_3)
    G_{\gamma\delta \leftarrow \alpha\beta}(t_2)
    G_{0\beta}(t_1),
    \label{eq:response-rephasing-SE}
\end{align}
where $G_{\gamma\delta \leftarrow \alpha\beta}(t_2)$ is the matrix element of the time-evolution operator defined by $\rho_{\gamma\delta}(t)=\sum_{\alpha\beta}G_{\gamma\delta \leftarrow \alpha\beta}(t-s)\rho_{\alpha\beta}(s)$ and $G_{\alpha\beta}(t)$ describes the time evolution of the $\lvert\alpha\rangle\langle\beta\rvert$ coherence. For simplicity, we set $\hbar = 1$ as follows.

For simplicity, we focus on the limit of the short entanglement time, $T_\e \to 0$.
This is reasonable when $T_\e$ is much shorter than the characteristic timescales of the dynamics under investigation.
Numerical simulations demonstrate that for a $\beta$-$\mathrm{BaB_2O_4}$ crystal, an entanglement time of $T_\e = 10\,{\rm fs}$ can be achieved for a crystal length of $L =0.056\,{\rm mm}$, making it possible to study real-time dynamics on ultrafast timescales approximate to subpicoseconds \cite{Fujihashi:2023pr}.
Given this scenario, the six-body correlation functions of the field in the rephasing SE signal are computed as
\begin{align}
    C_{\SE,\i\s}^{(\rephasing)}(\omega,t,\tau;s_3,s_2,s_1)
    &=
    \delta(t-s_3-\tau)
    \delta(t-s_3-s_2+\dt-\tau)
\notag \\
    &\quad\times     
    e^{-i\omega(t-s_3-s_2-s_1)}
    e^{-i\omega_\p(s_1+\dt)},
\end{align}
\begin{align}
    C_{\SE,\s\i}^{(\rephasing)}(\omega,t,t;s_3,s_2,s_1)
    &=
    \delta(-s_3-s_2-s_1+\dt)
\notag \\
    &\quad\times       
    e^{-i\omega(s_2+\dt)}
    e^{-i\omega_\p(s_1-\dt)}.
    \label{eq:correlation-function-SE}
\end{align}
Using Eqs.~\eqref{eq:response-rephasing-SE}--\eqref{eq:correlation-function-SE}, the rephasing contribution of the SE signal is obtained as
\begin{align}
    S_{\SE}^{(\rephasing)}(\omega;\dt)
    &=
    -
    \cos[(2\omega-\omega_\p)\dt]
    \sum_{\alpha\beta\gamma\delta}
    \mu_{\delta0}
    \mu_{\gamma0}
    \mu_{\beta0}
    \mu_{\alpha0}
\notag \\
    &\quad\times       
    G_{\gamma0}[\omega]
    G_{\gamma\delta \leftarrow \alpha\beta}(\dt)
    G_{0\beta}[\omega_\p - \omega],
    \label{eq:tpc-rephasing-SE-signal}
\end{align}
where $G_{\alpha\beta}[\omega] = \int_0^\infty dt\,e^{i\omega t} G_{\alpha\beta}(t)$.
Equation~\eqref{eq:tpc-rephasing-SE-signal} includes the oscillatory component depending on the values of $\omega$ and $\omega_\p$.
This is simply a phase shift owing to the introduction of the delay interval $\dt$, and does not reflect information about the molecular system, such as electronic coherence.
This undesirable oscillation can be eliminated by fixing $\omega=\omega_\p/2$.
In this case, Eq.~\eqref{eq:tpc-rephasing-SE-signal} temporally resolves the excited state dynamics of $\lvert e_\alpha\rangle\langle e_\beta \rvert \to \lvert e_\gamma \rangle\langle e_\delta \rvert$.

\subsection{Rephasing ESA contribution}

The six-body correlation function of the field in the rephasing ESA signal is computed as follows:
\begin{align}
    C_{\ESA,\i\s}^{(\rephasing)}&(\omega,t,\tau;s_3,s_2,s_1)    
\notag \\
    &=
    \langle
    \E^-(t-s_3-s_2-s_1) \E_{\s}^-(\tau) \E_{\s}^+(\tau) \E_{\i}^-(\omega)
\notag \\
    &\quad\times     
    \E^+(t-s_3) \E^+(t-s_3-s_2)
    \rangle
\notag \\
    &=
    \langle \psi_\twin \rvert 
    \E_{\s}^-(t-s_3-s_2-s_1)\E_{\s}^-(\tau)\E_{\s}^+(\tau)\E_{\i}^-(\omega)
\notag \\
    &\quad\times       
    \E_{\i}^+(t-s_3+\dt)\E_{\s}^+(t-s_3-s_2)
    \lvert\psi_\twin \rangle
\notag \\
    &\quad+
    \langle \psi_\twin \rvert 
    \E_{\s}^-(t-s_3-s_2-s_1)\E_{\s}^-(\tau)\E_{\s}^+(\tau)\E_{\i}^-(\omega)
\notag \\
    &\quad\times      
    \E_{\s}^+(t-s_3)\E_{\i}^+(t-s_3-s_2+\dt)
    \lvert\psi_\twin \rangle.
    \label{eq:correlation-function-ESA}
\end{align}
Because the operator $\E_{\s}^+$, i.e., the annihilation operator of the signal photon $\hat{a}_\s$ acts on the twin-photon state $\lvert\psi_\twin \rangle$ twice in succession, Eq.~\eqref{eq:correlation-function-ESA} becomes
\begin{align}
    C_{\ESA,\i\s}^{(\rephasing)}&(\omega,t,\tau;s_3,s_2,s_1)    
    =0.
\end{align}
Similarly, the correlation function of the field in the non-rephasing ESA signal vanishes. 
In contrast to entangled two-photon spectroscopy \cite{Ishizaki:2020jl} and conventional 2D Fourier-transformed photon-echo spectra \cite{Khalil:2003fn}, 
two-photon coincidence entangled two-photon spectroscopy therefore does not produce signals that originate from the ESA pathway.

Noteworthy, is that this advantage is owing to the photon number correlations, not the time-frequency correlations between the twin photons. The time-frequency correlations are only used to enable time-resolved spectroscopy with monochromatic pumping \cite{Ishizaki:2020jl}.

\subsection{Total signal}

By calculating other signal contributions as well as rephasing SE and ESA, Eq.~\eqref{eq:tpc-signal} is expressed as
\begin{align}
    S_\mathrm{TPC}(\omega;\dt)
    =
    \Re \,
    \left[
    S_{\SE}(\omega;\dt)
    +
    S_{\GSB}(\omega;\dt)
    +
    S_\mathrm{c}(\omega)
    \right],
    \label{eq:tpc-total-signal}
\end{align}
with the SE and GSB contributions,
\begin{align}
    S_{\SE}(\omega;\dt)
    &=
    -
    \cos[(2\omega-\omega_\p)\dt]
    \sum_{\alpha\beta\gamma\delta}
    \mu_{\delta0}
    \mu_{\gamma0}
    \mu_{\beta0}
    \mu_{\alpha0}
\notag \\
    &\quad\times         
    G_{\gamma0}[\omega]
    G_{\gamma\delta \leftarrow \beta\alpha}(\dt)
\notag \\
    &\quad\times      
    \left(
    G_{\alpha0}[\omega_\p - \omega]
    +
    G_{\beta0}^\ast[\omega_\p - \omega]
    \right),
\end{align}
\begin{align}
    S_{\GSB}(\omega;\dt)
    &=
    -
    \sum_{\alpha\beta}
    \mu_{\beta0}^2
    \mu_{\alpha0}^2
    G_{\beta0}[\omega]
    G_{00 \leftarrow 00}(\dt)
\notag \\
    &\quad\times      
    \left(
    G_{\alpha0}[\omega_\p - \omega]
    +
    G_{\alpha0}^\ast[\omega_\p - \omega]
    \right)
\notag \\
    &\quad-
    e^{i(2\omega-\omega_\p)\dt}
    \sum_{\alpha\beta}
    \mu_{\beta0}^2
    \mu_{\alpha0}^2
    G_{\beta0}[\omega]
    G_{00 \leftarrow 00}(\dt)
\notag \\
    &\quad\times       
    G_{\alpha0}[\omega_\p - \omega],
\end{align}
where the last term in Eq.~\eqref{eq:tpc-total-signal} originates from the field commutator, and is written as
\begin{align}
    S_\mathrm{c}(\omega)
    &=
    -
    \sum_{\alpha\beta}
    \mu_{\beta0}^2
    \mu_{\alpha0}^2
    \left(
    G_{\beta0}[\omega]
    G_{\alpha0}[\omega_\p - \omega]
    \right.
\notag \\
    &\quad
    \left.    
    +
    G_{\beta0}[\omega_\p -\omega]
    G_{00 \leftarrow 00}[\omega_\p -\omega]
    \right).
    \label{eq:tpc-dt-independent-term} 
\end{align}
In deriving Eq.~\eqref{eq:tpc-total-signal}, we employed the approximation of $G_{\gamma\delta \leftarrow \alpha\beta}(\dt-s_1)G_{\alpha0}(s_1)\simeq G_{\gamma\delta \leftarrow \alpha\beta}(\dt)G_{\alpha0}(s_1)$ for the non-rephasing Liouville pathways \cite{Cervetto:2004gm}.
This approximation is justified when the response function varies slowly as a function of the waiting time, $s_2$.
As mentioned in Sec.~3.1, fixing $\omega=\omega_\p/2$, Eq.~\eqref{eq:tpc-total-signal} reduces to
\begin{align}
    S_{\SE}(\omega=\omega_\p/2;\dt)
    &=
    -
    \sum_{\alpha\beta\gamma\delta}
    \mu_{\delta0}
    \mu_{\gamma0}
    \mu_{\beta0}
    \mu_{\alpha0}
    G_{\gamma0}[\omega_\p/2]
    G_{\gamma\delta \leftarrow \alpha\beta}(\dt)
\notag \\
    &\quad\times        
    \left(
    G_{\alpha0}[\omega_\p/2]
    +
    G_{\beta0}^\ast[\omega_\p/2]
    \right),
    \label{eq:tpc-SE-signal}
\end{align}
\begin{align}
    S_{\GSB}(\omega=\omega_\p/2;\dt)
    &=
    -
    \sum_{\alpha\beta}
    \mu_{\beta0}^2
    \mu_{\alpha0}^2
    G_{\beta0}[\omega_\p/2]
    G_{00 \leftarrow 00}(\dt)
\notag \\
    &\quad\times          
    \left(
    2G_{\alpha0}[\omega_\p/2]
    +
    G_{\alpha0}^\ast[\omega_\p/2]
    \right).
    \label{eq:tpc-GSB-signal}
\end{align}
Moreover, to remove the $\dt$-independent contributions, the difference spectrum is considered, $\Delta S(\omega;\dt)=S(\omega;\dt)-S(\omega;\dt=0)$, which contains only the SE contribution as a function of $\dt$.
Thus, Eq.~\eqref{eq:tpc-SE-signal} shows that two-photon coincidence entangled two-photon spectroscopy is able to obtain spectroscopic information on the excited state dynamics by sweeping the pump frequency $\omega_\p$.

\section{Numerical results}

Next, we compare the proposed two-photon coincidence entangled two-photon spectroscopy with entangled two-photon spectroscopy for a model system.

We consider the electronic excitations in a coupled-trimer.
The molecular Hamiltonian is given by $\hat{H}_{\rm mol} = \hat{H}_{\rm ex} + \hat{H}_{\rm ex-env} + \hat{H}_{\rm env}$ \cite{Fujihashi:2015kz}: 
These three terms represent the electronic excitation Hamiltonian, the environmental Hamiltonian, their interaction, respectively.
The electronic excitation Hamiltonian is expressed as $\hat{H}_{\rm ex}=\sum_m \hbar \Omega_m \hat{B}^\dagger_m \hat{B}_m + \sum_{m \neq n} \hbar J_{mn} \hat{B}^\dagger_m \hat{B}_n$, where $\hbar\Omega_m$ is the Franck--Condon transition energy of the $m$th pigment, $\hbar J_{mn}$ is the electronic coupling between pigments, and the excitation creation operator $\hat{B}^\dagger_m$ is introduced for the excitation vacuum $\vert 0 \rangle$ such that $\vert m \rangle = \hat{B}^\dagger_m \vert 0 \rangle$ and $\vert mn \rangle = \hat{B}^\dagger_m \hat{B}^\dagger_n \vert 0 \rangle$.
In the eigenstate representation, the excitation Hamiltonian can be expressed as $\hat{H}_{\rm ex}=\epsilon_0  \lvert 0 \rangle\langle 0 \rvert + \sum_\alpha \epsilon_\alpha \lvert e_\alpha \rangle\langle e_\alpha \rvert +\sum_{\bar\gamma} \epsilon_{\bar\gamma} \lvert  f_{\bar\gamma} \rangle\langle f_{\bar\gamma} \rvert$, where $\lvert e_\alpha \rangle=\sum_m V_{m\alpha} \hat{B}^\dagger_m \vert 0 \rangle$ and $\vert f_{\bar\gamma} \rangle = \sum_{mn} W_{mn,{\bar\gamma}} \hat{B}^\dagger_m \hat{B}^\dagger_n \vert 0 \rangle$.
Accordingly, the exciton transition dipole moments are expressed as $\mu_{\alpha0}=\sum_m V_{\alpha m}^{-1} \mu_{m0}$ and $\mu_{\bar\gamma \alpha}= \sum_{mn} W_{{\bar\gamma}(mn)}^{-1}V_{\alpha m}^{-1} \mu_{n0}$.
In this study, the environment is not treated explicitly, albeit the time evolution operator is described phenomenologically as $G_{\alpha\beta}(t)=e^{-(i\omega_{\alpha\beta}+\Gamma_{\rm env})t}$, where $\hbar\omega_{\alpha\beta}=\epsilon_\alpha - \epsilon_\beta$ represents the energy gap between excitons.

For numerical calculations, we set the Franck--Condon transition energies of pigments 1, 2, and 3 to $\hbar\Omega_1 =12500\,{\rm cm}^{-1}$, $\hbar\Omega_2 =12400\,{\rm cm}^{-1}$, and $\hbar\Omega_3 =12200\,{\rm cm}^{-1}$, respectively.
The parameters are set to $\hbar J_{12} =-20\,{\rm cm}^{-1}$, $\hbar J_{23} =20\,{\rm cm}^{-1}$, $\hbar J_{31} =30\,{\rm cm}^{-1}$, and $\Gamma_{\rm env}^{-1}=50\,{\rm fs}$.
For simplicity, we set the transition dipole strengths to $\mu_{10}=\mu_{20}=\mu_{30}=1$.

For comparison, we consider the 2D spectrum obtained with the entangled two-photon spectroscopy in Ref.~\citenum{Ishizaki:2020jl}, which is expressed as
\begin{align}
    S(\omega;\dt)
    =
    \Re \,
    \left[
    S_{\SE}(\omega;\dt)
    +
    S_{\GSB}(\omega;\dt)
    +
    S_{\ESA}(\omega;\dt)
    \right],
    \label{eq:two-photon-spectroscopy}  
\end{align}    
with the SE, GSB, and ESA contributions,
\begin{align}
    S_{\SE}(\omega;\dt)
    &=
    -
    \sum_{\alpha\beta\gamma\delta}
    \mu_{\delta0}
    \mu_{\gamma0}
    \mu_{\beta0}
    \mu_{\alpha0}
    G_{\gamma0}[\omega]  
    G_{\gamma\delta \leftarrow \beta\alpha}(\dt)  
\notag \\
    &\quad\times        
    \left(
    G_{\alpha0}[\omega_\p - \omega]
    +
    G_{\beta0}^\ast[\omega_\p - \omega]
    \right),
\end{align}
\begin{align}
    S_{\GSB}(\omega;\dt)
    &=
    -
    \sum_{\alpha\beta}
    \mu_{\beta0}^2
    \mu_{\alpha0}^2
    G_{\beta0}[\omega]
    G_{00 \leftarrow 00}(\dt)
\notag \\
    &\quad\times      
    \left(
    G_{\alpha0}[\omega_\p - \omega]
    +
    G_{\alpha0}^\ast[\omega_\p - \omega]
    \right),
\end{align}
\begin{align}
    S_{\ESA}(\omega;\dt)
    &=
    -
    \sum_{\alpha\beta\gamma\delta\bar\epsilon}
    \mu_{\bar\epsilon\delta}
    \mu_{\bar\epsilon\gamma}
    \mu_{\beta0}
    \mu_{\alpha0}
    G_{\bar\epsilon\delta}[\omega]  
    G_{\gamma\delta \leftarrow \beta\alpha}(\dt)  
\notag \\
    &\quad\times        
    \left(
    G_{\alpha0}[\omega_\p - \omega]
    +
    G_{\beta0}^\ast[\omega_\p - \omega]
    \right).
\end{align}
Note that the $\dt$-independent term is ignored for simplicity. Equation~\eqref{eq:two-photon-spectroscopy} provides identical information contents as that of 2D Fourier-transformed photon-echo spectra with classical laser pulses \cite{Ishizaki:2020jl}.

\begin{figure}
    \centering
    \includegraphics{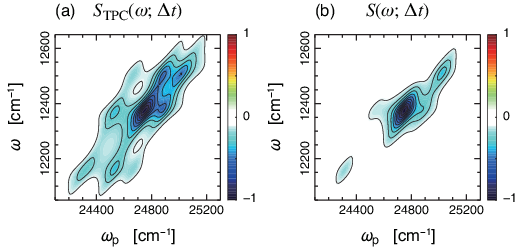}
    \caption{Comparison of the proposed two-photon coincidence entangled two-photon spectroscopy with the entangled two-photon spectroscopy. (a) Signal, $S_\mathrm{TPC}(\omega;\dt)$, in Eq.~\eqref{eq:tpc-total-signal} except for the $\dt$-independent term in Eq.~\eqref{eq:tpc-dt-independent-term}. (b) Signal, $S(\omega;\dt)$, in Eq.~\eqref{eq:two-photon-spectroscopy}. Delay time is $\dt=0$. Signals are computed for a coupled-trimer.
    Normalization of contour plots~(a) and (b) is such that the maximum value of each spectrum is unity, and equally spaced contour levels ($\pm 0.1$, $\pm 0.2$, \dots) are drawn.
    }
    \label{fig2}
\end{figure}

Figures~\ref{fig2}(a) and (b) present the signal, $S_\mathrm{TPC}(\omega;\dt)$, in Eq.~\eqref{eq:tpc-total-signal}, and the signal, $S(\omega;\dt)$, in Eq.~\eqref{eq:two-photon-spectroscopy}, respectively. 
The delay time is $\dt=0$. 
Although nine peaks should appear in the coupled-trimer 2D spectrum, only diagonal peaks are observed in the spectrum in Fig.~\ref{fig2}(b). 
This is because the SE and GSB signals cancel out the ESA signal that appears near them.
In contrast, for the two-photon coincidence entangled two-photon spectroscopy approach, the ESA signal is eliminated owing to the photon number correlations, resulting in clear observation of all nine peaks, as shown in Fig.~\ref{fig2}(a).
Hence, the proposed spectroscopy can help to ease the spectral interpretation in complex molecular systems comprising multiple molecules.

For finite delay times, $\dt>0$, however, the Eq.~\eqref{eq:tpc-total-signal} signal includes the oscillatory component depending on the values of $\omega$ and $\omega_\p$.
As shown in Eqs.~\eqref{eq:tpc-SE-signal} and \eqref{eq:tpc-GSB-signal}, this undesired oscillatory component can be eliminated by fixing $\omega=\omega_\p/2$, however, in such a case only spectral information along the diagonal line is obtained. Therefore, future research will concentrate on investigating this problem.

\section{Discussion}
In the previous section, we demonstrated that the selective elimination of the ESA signal can be achieved by coincidence counting the entangled photon pairs. 
Next, we determine the need for the photon number correlations to achieve pathway selectivity. For this purpose, we consider the initial state of the electric field as composed of two coherent pulses
\begin{align}
    \lvert\psi_\text{coh} \rangle
    =
    \lvert C_\text{pu} \rangle
    \lvert C_\text{pr} \rangle,
\end{align}
where $\lvert C_\sigma \rangle$ represents a multimode coherent state defined by \cite{loudon2000quantum}
\begin{align}
    \lvert C_\sigma \rangle
    &=
    \prod_\omega
    e^{-|\alpha_\sigma(\omega)|^2 /2}
    \sum_{n=0}
    \frac{\alpha_\sigma (\omega)^n}{\sqrt{n!}}
    \hat{a}_\sigma^\dagger(\omega)^n
    \lvert \vac \rangle.
    \label{eq:coherent-state}
\end{align}
In the above equation, $\alpha_\text{pu} (\omega)$ and $\alpha_\text{pr} (\omega)$ are the normalized spectral envelope of the pump and probe laser pulses, respectively. 
The positive-frequency component of the field operator is expressed as $\E^+(t) = \E_\text{pu}^+(t) + \E_\text{pr}^+(t+\dt)$, where $\E_\sigma^+(t)=(2\pi)^{-1}\int d\omega\,\hat{a}_\sigma(\omega)e^{-i\omega t}$.
The correlation function of the field operator in the rephasing ESA signal can be calculated as follows:
\begin{align}
    C_{\ESA}^{(\rephasing)}&(\omega,t,\tau;s_3,s_2,s_1)  
\notag \\
    &=
    \langle
    \E^-(t-s_3-s_2-s_1)\E_\text{pu}^-(\tau)\E_\text{pu}^+(\tau)\E_\text{pr}^-(\omega)
\notag \\
    &\quad\times      
    \E^+(t-s_3)\E^+(t-s_3-s_2)
    \rangle
\notag \\
    &=
    \langle \psi_\text{coh} \rvert 
    \E_\text{pu}^-(t-s_3-s_2-s_1)\E_\text{pu}^-(\tau)\E_\text{pu}^+(\tau)\E_\text{pr}^-(\omega)
\notag \\
    &\quad\times        
    \E_\text{pr}^+(t-s_3+\dt)\E_\text{pu}^+(t-s_3-s_2)
    \lvert\psi_\text{coh} \rangle
\notag \\
    &\quad
    +
    \langle \psi_\text{coh} \rvert 
    \E_\text{pu}^-(t-s_3-s_2-s_1)\E_\text{pu}^-(\tau)\E_\text{pu}^+(\tau)\E_\text{pr}^-(\omega)
\notag \\
    &\quad\times       
    \E_\text{pu}^+(t-s_3)\E_\text{pr}^+(t-s_3-s_2+\dt)
    \lvert\psi_\text{coh} \rangle
\notag \\
    &=
    \langle C_\text{pu} \rvert 
    \E_\text{pu}^-(t-s_3-s_2-s_1)\E_\text{pu}^-(\tau)\E_\text{pu}^+(\tau)
\notag \\
    &\quad\times         
    \E_\text{pu}^+(t-s_3-s_2)
    \lvert C_\text{pu} \rangle
\notag \\
    &\quad\times        
    \langle C_\text{pr} \rvert 
    \E_\text{pr}^-(\omega)\E_\text{pr}^+(t-s_3+\dt)
    \lvert C_\text{pr} \rangle
\notag \\
    &\quad+
    \langle C_\text{pu} \rvert 
    \E_\text{pu}^-(t-s_3-s_2-s_1)\E_\text{pu}^-(\tau)\E_\text{pu}^+(\tau)
    \E_\text{pu}^+(t-s_3)
    \lvert C_\text{pu} \rangle
\notag \\
    &\quad\times    
    \langle C_\text{pr} \rvert 
    \E_\text{pr}^-(\omega)\E_\text{pr}^+(t-s_3-s_2+\dt)
    \lvert C_\text{pr} \rangle,
\end{align}
We have factorized the terms corresponding to the operators for the pump and probe fields, respectively. 
By using Eq.~\eqref{eq:coherent-state}, we show that
\begin{multline}
    \langle C_\text{pu} \rvert 
    \E_\text{pu}^-(t-s_3-s_2-s_1)
    \E_\text{pu}^-(\tau) 
    \E_\text{pu}^+(\tau)
    \E_\text{pu}^+(t-s_3-s_2)
    \lvert C_\text{pu} \rangle
\\
    =
    \alpha_\text{pu}^\ast(t-s_3-s_2-s_1)
    \alpha_\text{pu}^\ast(\tau)
    \alpha_\text{pu}(\tau)
    \alpha_\text{pu}(t-s_3-s_2),
\end{multline}
where $\alpha_\text{pu}(t)=(2\pi)^{-1}\int d\omega\, \alpha_\text{pu}(\omega)e^{-i\omega t}$.
Similarly, the correlation function of the operators for the probe field is obtained as a product of $\alpha_\text{pr}(t)$.
 This result demonstrates that the rephasing ESA signal survives when coherent light is used. A similar calculation can be used to confirm that the correlation function of the field operator in the non-rephasing ESA signal is nonzero. 
 Therefore, the selective elimination of the ESA signal achieved by the coincidence counting of the entangled photon pair cannot be emulated by classical coherent light.

\section{Conclusion}
In this work, we theoretically proposed time-resolved spectroscopy based on the coincidence counting of the entangled photon pair. We demonstrated that selective elimination of the ESA signal is possible by using coincidence counting of entangled photon pairs, simplifying spectral interpretation. This selective elimination cannot be emulated by classical coherent light.
We believe that the results of this study provide insight into the design of applications of quantum light for spectroscopy that offer actual quantum advantages.

In this study, our investigation was limited to the weak down-conversion regime. However, detecting the nonlinear optical response from just one pair of entangled photons will be considerably challenging owing to its extremely weak nature. Consequently, investigating the feasibility of two-photon coincidence entangled two-photon spectroscopy in the high-gain regime for practical utilization is critical. As Ref.~\citenum{Fujihashi:2023pr} theoretically illustrates, when the entanglement time is sufficiently short compared with the characteristic timescales of the dynamics under investigation, entangled two-photon spectroscopy, even in the high gain regime is capable of temporally resolving the excitation dynamics. 
Therefore, future work will focus on expanding this research direction.

\begin{acknowledgments}
This study was supported by the MEXT Quantum Leap Flagship Program (Grant Number~JPMXS0118069242) and JSPS KAKENHI (Grant Number~JP21H01052).
Y.F. acknowledges support from JSPS KAKENHI (Grant Number~JP23K03341).
\end{acknowledgments}

\section*{Data Availability Statement}
The data that support the findings of this study are available from the corresponding author upon reasonable request.


\appendix
\section{\label{app:secA}The twin photon state}

The photon pair produced by a type-II PDC process can be described by the wave function \cite{Grice:1997ht}:
\begin{align}
	\lvert \psi_\twin \rangle
	=
	\int d\omega_1 \int d\omega_2
    \,
	f(\omega_1,\omega_2)
	\adagger_\s(\omega_1) \adagger_\i(\omega_2)
	\vert \vac \rangle,
	\label{eq:photon-state}
\end{align}
where $\hat{a}_\s^\dagger(\omega)$ and $\hat{a}_\i^\dagger(\omega)$ are the creation operators of the signal and idler photon, respectively. 
The two-photon amplitude, $f(\omega_1,\omega_2)$ is expressed as $f(\omega_1,\omega_2) = \zeta \alpha_\p(\omega_1+\omega_2) \sinc[\Delta k(\omega_1,\omega_2)L/2]$, where $\alpha_\p(\omega)$ is the normalized pump envelope, $L$ is the length of a nonlinear crystal, and the sinc function originates from phase-matching.
Expanding the wave vector mismatch $\Delta k(\omega_1,\omega_2)$ to first order in the frequencies $\omega_1$ and $\omega_2$ around the centre frequencies of the generated beams, $\bar\omega_\s$ and $\bar\omega_\i$, we obtain $\Delta k(\omega_1,\omega_2) = (\omega_1 - \bar\omega_\s)T_\s + (\omega_2 - \bar\omega_\i)T_\i$ with $T_\lambda=(v_\p^{-1} - v_\lambda^{-1})L$ \cite{Rubin:1994ed,Keller:1997hj}, where $v_\p$ and $v_\lambda$ are the group velocities of the pump laser and one of the generated beams at central frequency $\bar\omega_\lambda$, respectively. 
All other constants are merged into factor $\zeta$, which corresponds to the conversion efficiency of the PDC process.

In this study, we address the case of monochromatic pumping $\alpha_\p(\omega_1+\omega_2)=\delta(\omega_1+\omega_2-\omega_\p)$. 
Thus, the two-photon amplitude is recast as 
\begin{align}
    f(\omega_1,\omega_2)
    =
    \zeta 
    \delta(\omega_1 + \omega_2 - \omega_\p) \,
    \sinc\frac{(\omega_2 - \bar\omega_\i) T_\e}{2}.
    \label{eq:two-photon-amp}
\end{align}
The so-called entanglement time $T_\e= 
\lvert
T_\s-T_\i
\rvert
$ is the maximum time difference between twin photons leaving the crystal \cite{Saleh:1998vl}.

\section{\label{app:secB}Six-body correlation function of the electric field operators}

In the limit of short entanglement time, $T_\e \to 0$, the six-body correlation functions in Eq.~\eqref{eq:tpc-signal-contribution} are computed as follows:
\begin{align}
    C_{\SE,\i\s}^{(\rephasing)}&(\omega,t,\tau;s_3,s_2,s_1)
\notag \\    
    &=
    \langle
    \E^-(t-s_3-s_2-s_1)\E^+(t-s_3)\E_{\s}^-(\tau)\E_{\s}^+(\tau)\E_{\i}^-(\omega)
\notag \\
    &\quad\times          
    \E^+(t-s_3-s_2)
    \rangle
\notag \\
    &=
    \delta(t-s_3-\tau)
    \delta(t-s_3-s_2+\dt-\tau)
\notag \\
    &\quad\times        
    e^{-i\omega(t-s_3-s_2-s_1)}
    e^{-i\omega_\p(s_1+\dt)},
\end{align}
\begin{align}
    C_{\SE,\s\i}^{(\rephasing)}&(\omega,t,t;s_3,s_2,s_1)
\notag \\    
    &=
    \langle
    \E^-(t-s_3-s_2-s_1)\E^+(t-s_3)\E_{\i}^-(\omega)\E_{\i}^+(\omega)\E_{\s}^-(t)
\notag \\
    &\quad\times        
    \E^+(t-s_3-s_2)
    \rangle
\notag \\
    &=
    \delta(-s_3-s_2-s_1+\dt)
    e^{-i\omega(s_2+\dt)}
    e^{-i\omega_\p(s_1-\dt)},
\end{align}
\begin{align}
    C_{\SE,\i\s}^{(\nonrephasing)}&(\omega,t,\tau;s_3,s_2,s_1)
\notag \\    
    &=
    \langle
    \E^-(t-s_3-s_2)\E^+(t-s_3)\E_{\s}^-(\tau)\E_{\s}^+(\tau)\E_{\i}^-(\omega)
\notag \\
    &\quad\times          
    \E^+(t-s_3-s_2-s_1)
    \rangle
\notag \\
    &=
    \delta(t-s_3-\tau)
    \delta(t-s_3-s_2-s_1+\dt-\tau)
\notag \\
    &\quad\times        
    e^{-i\omega(t-s_3-s_2)}
    e^{-i\omega_\p(-s_1+\dt)},
\end{align}
\begin{align}
    C_{\SE,\s\i}^{(\nonrephasing)}&(\omega,t,t;s_3,s_2,s_1)
\notag \\    
    &=
    \langle
    \E^-(t-s_3-s_2)\E^+(t-s_3)\E_{\i}^-(\omega)\E_{\i}^+(\omega)\E_{\s}^-(t)
\notag \\
    &\quad\times        
    \E^+(t-s_3-s_2-s_1)
    \rangle
\notag \\
    &=
    \delta(s_3+s_2-\dt)
    e^{-i\omega(s_2+s_1+\dt)}
    e^{i\omega_\p(s_1+\dt)},
\end{align}
\begin{align}
    C_{\GSB,\i\s}^{(\rephasing)}(&\omega,t,\tau;s_3,s_2,s_1)
\notag \\    
    &=
    \langle
    \E^-(t-s_3-s_2-s_1)\E^+(t-s_3-s_2)\E_{\s}^-(\tau)\E_{\s}^+(\tau)\E_{\i}^-(\omega)
\notag \\
    &\quad\times          
    \E^+(t-s_3)
    \rangle
\notag \\
    &=
    \delta(t-s_3-s_2-\tau)
    \delta(t-s_3+\dt-\tau)
\notag \\
    &\quad\times          
    e^{-i\omega(t-s_3-s_2-s_1)}
    e^{-i\omega_\p(s_2+s_1+\dt)},
\end{align}    
\begin{align}
    C_{\GSB,\s\i}^{(\rephasing)}&(\omega,t,t;s_3,s_2,s_1)
\notag \\    
    &=
    \langle
    \E^-(t-s_3-s_2-s_1)\E^+(t-s_3-s_2)\E_{\i}^-(\omega)\E_{\i}^+(\omega)\E_{\s}^-(t)
\notag \\
    &\quad\times         
    \E^+(t-s_3)
    \rangle
\notag \\    
    &=
    \delta(-s_3-s_2-s_1+\dt)
    e^{-i\omega(-s_2+\dt)}
    e^{i\omega_\p(-s_2-s_1+\dt)},
\end{align}    
\begin{align}
    C_{\GSB,\i\s}^{(\nonrephasing)}(&\omega,t,\tau;s_3,s_2,s_1)
\notag \\    
    &=
    \langle
    \E_{\s}^-(\tau)\E_{\s}^+(\tau)\E_{\i}^-(\omega)
    \E^+(t-s_3)\E^-(t-s_3-s_2)
\notag \\
    &\quad\times         
    \E^+(t-s_3-s_2-s_1)
    \rangle
\notag \\
    &=
    \delta(t-s_3-s_2-\tau)
    \delta(s_2+s_1+\dt)
    e^{-i\omega\tau}   
    e^{-i\omega_\p(t-s_3+\dt-\tau)}
\notag \\
    &\quad    
    +
    \delta(t-s_3-s_2-\tau)
    \delta(s_2+s_1-\dt)
    e^{-i\omega\tau}
    e^{-i\omega_\p(t-s_3-\tau)}
\notag \\
    &\quad    
    +
    \delta(t-s_3-s_2-s_1+\dt-\tau)
    \delta(s_2)
    e^{-i\omega\tau},
\end{align}    
\begin{align}
    C_{\GSB,\s\i}^{(\nonrephasing)}&(\omega,t,t;s_3,s_2,s_1)
\notag \\    
    &=
    \langle
    \E_{\i}^-(\omega)\E_{\i}^+(\omega)\E_{\s}^-(t)
    \E^+(t-s_3)\E^-(t-s_3-s_2)
\notag \\
    &\quad\times         
    \E^+(t-s_3-s_2-s_1)
    \rangle
\notag \\    
    &=
    \delta(s_2+s_1-\dt)
    e^{-i\omega(s_3+s_2-\dt)}
    e^{i\omega_\p s/3}
\notag \\
    &\quad        
    +
    \delta(s_2+s_1+\dt)
    e^{-i\omega(s_3+s_2-\dt)}
    e^{i\omega_\p (s/3-\dt)}
\notag \\
    &\quad    
    +
    \delta(s_1)
    e^{-i(\omega-\omega_\p)(s_3+s_2+s_1)},
\end{align}    
\begin{align}
    C_{\ESA,\i\s}^{(\rephasing)}&(\omega,t,\tau;s_3,s_2,s_1)
\notag \\    
    &=
    \langle
    \E^-(t-s_3-s_2-s_1)\E_{\s}^-(\tau)\E_{\s}^+(\tau)\E_{\i}^-(\omega)
\notag \\
    &\quad\times      
    \E^+(t-s_3)\E^+(t-s_3-s_2)
    \rangle
\notag \\
    &=0,
\end{align}
\begin{align}
    C_{\ESA,\s\i}^{(\rephasing)}&(\omega,t,t;s_3,s_2,s_1)
\notag \\    
    &=
    \langle
    \E^-(t-s_3-s_2-s_1)\E_{\i}^-(\omega)\E_{\i}^+(\omega)\E_{\s}^-(t)
\notag \\
    &\quad\times      
    \E^+(t-s_3)\E^+(t-s_3-s_2)
    \rangle
\notag \\
    &=0,
\end{align}
\begin{align}
    C_{\ESA,\i\s}^{(\nonrephasing)}&(\omega,t,\tau;s_3,s_2,s_1)
\notag \\    
    &=
    \langle
    \E^-(t-s_3-s_2)\E_{\s}^-(\tau)\E_{\s}^+(\tau)\E_{\i}^-(\omega)
\notag \\
    &\quad\times      
    \E^+(t-s_3)\E^+(t-s_3-s_2-s_1)
    \rangle
\notag \\
    &=0,
\end{align}
and
\begin{align}
    C_{\ESA,\s\i}^{(\nonrephasing)}&(\omega,t,t;s_3,s_2,s_1)
\notag \\    
    &=
    \langle
    \E^-(t-s_3-s_2)\E_{\i}^-(\omega)\E_{\i}^+(\omega)\E_{\s}^-(t)
\notag \\
    &\quad\times      
    \E^+(t-s_3)\E^+(t-s_3-s_2-s_1)
    \rangle
\notag \\
    &=0.
\end{align}




%

\end{document}